\documentclass[aps,prx,twocolumn,superscriptaddress,floatfix,nofootinbib,longbibliography]{revtex4-1}

\usepackage{mathrsfs}
\usepackage{amssymb}
\usepackage{amsmath}
\usepackage{amstext}
\usepackage[english]{babel}
\usepackage{graphicx}
\usepackage{bm}
\usepackage{float}
\usepackage{color}
\usepackage[breaklinks]{hyperref}
\hypersetup{colorlinks=true, linkcolor=blue, citecolor=blue, filecolor=blue, urlcolor=blue}

\makeatletter

\newcommand{\Rmnum}[1]{\expandafter\@slowromancap\romannumeral #1@}
\makeatother

\begin{document}

\title{Scrambling via Braiding of Nonabelions}

\author{Zhi-Cheng Yang}

\affiliation{Physics Department, Boston University, Boston,
  Massachusetts 02215, USA}
  
\author{Konstantinos Meichanetzidis}

\affiliation{School of Physics and Astronomy, University of Leeds, Leeds LS2 9JT, United Kingdom}

\author{Stefanos Kourtis}

\affiliation{Physics Department, Boston University, Boston,
  Massachusetts 02215, USA}

\author{Claudio Chamon}

\affiliation{Physics Department, Boston University, Boston,
  Massachusetts 02215, USA}

\date{\today}

\begin{abstract}
We study how quantum states are scrambled via braiding in systems of non-Abelian anyons through the lens of entanglement spectrum statistics. In particular, we focus on the degree of scrambling, defined as the randomness produced by braiding, at the same amount of entanglement entropy. To quantify the degree of randomness, we define a distance between the entanglement spectrum level spacing distribution of a state evolved under random braids and that of a Haar-random state, using the Kullback-Leibler divergence $D_{\mathrm{KL}}$. 
We study $D_{\mathrm{KL}}$ numerically for random braids of Majorana fermions (supplemented with random local four-body interactions) and Fibonacci anyons. For comparison, we also obtain $D_{\mathrm{KL}}$ for the Sachdev-Ye-Kitaev model of Majorana fermions with all-to-all interactions, random unitary circuits built out of (a) Hadamard (H), $\pi/8$ (T), and CNOT gates, and (b) random unitary circuits built out of two-qubit Haar-random unitaries. To compare the degree of randomness that different systems produce beyond entanglement entropy, we look at $D_{\mathrm{KL}}$ as a function of the Page limit-normalized entanglement entropy $S/S_{\mathrm{max}}$. Our results reveal a hierarchy of scrambling among various models --- even for the same amount of entanglement entropy --- at intermediate times, whereas all models exhibit the same late-time behavior. In particular, we find that braiding of Fibonacci anyons randomizes initial product states more efficiently than the universal H+T+CNOT set.

\end{abstract}

\maketitle


\section{introduction}
The notion of many-body quantum chaos plays a central role in understanding the emergence of statistical mechanical descriptions and thermodynamics of quantum systems under unitary time evolution (see for example~\cite{lindblad, gutzwiller, haake, ullmo, anatoli}). A precise quantitative formulation of quantum many-body chaos, in particular, remains an important problem. One possible diagnostic of chaos that has attracted recent interest is the ``out-of-time-ordered'' correlator (OTOC)~\cite{larkin, maldacena}, which generalizes the classical butterfly effect and Lyapunov exponent to quantum systems~\cite{shenker, shenker2, roberts, kitaev}. Indeed, the exponential growth behavior of the OTOC with a corresponding quantum Lyapunov exponent $\lambda_L$ has been confirmed in several large-$N$ theories (including large-$N$ gauge theories, as well as theories holographically dual to black holes~\cite{shenker, shenker2, roberts, kitaev, maldacena2, roberts2, stanford, chowdhury}), and weakly interacting disordered systems~\cite{patel}. 

A notion intimately related to chaos is \textit{scrambling}, which refers to the phenomenon that initially localized information of a system becomes undetectable under its own dynamics without measuring a significant fraction of all degrees of freedom~\cite{lashkari}. It was shown that the butterfly effect necessarily implies scrambling in quantum systems~\cite{hosur}. Scrambling can be captured by local entropy production under time evolution in chaotic systems~\cite{jonay}. Remarkably, the entanglement growth under random unitary dynamics belongs to the same universality class of Kardar-Parisi-Zhang equation for classical surface growth~\cite{nahum, kardar}. One may be tempted then to characterize the  degree of scrambling for different systems in terms of the growth rates of entanglement. However, the entanglement growth can behave in the same fashion in systems that are not truly chaotic, for example, under random Clifford dynamics~\cite{nahum, shaffer}.

\begin{figure}[t]
\centering
\includegraphics[width=.42\textwidth]{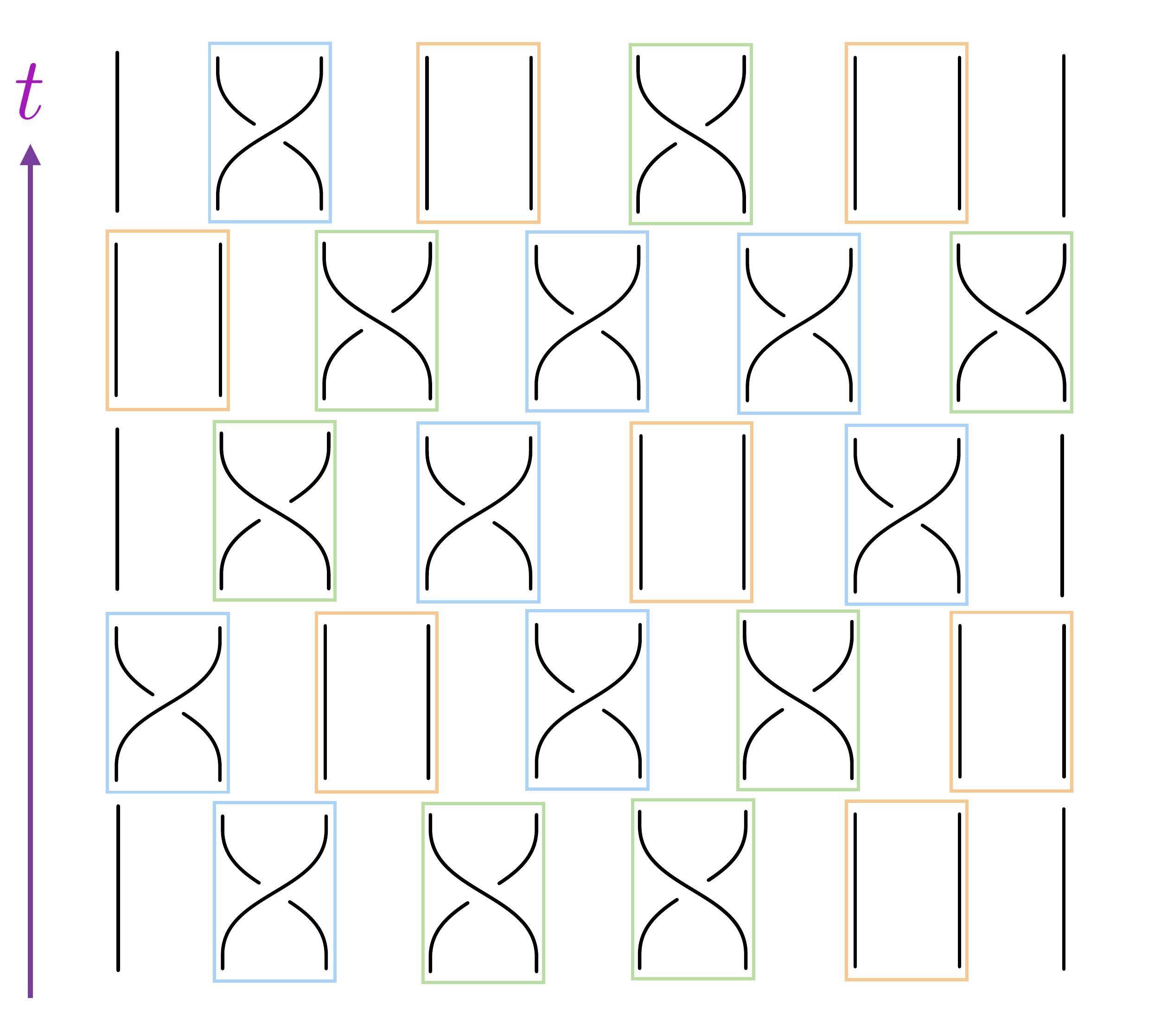}
\caption{(Color online) Depiction of braidings as a brick wall lattice of tiles representing elementary strand exchanges: 1 (no exchange), $T$ (overpass), and $T^{-1}$ (underpass) respectively. A random braid corresponds to a random choice of these tilings. }
\label{fig:braid_circuit} 
\end{figure}

Indeed, scrambling can exhibit different complexities depending on the randomness it produces, and there is a gap between maximal entanglement entropy and complete randomization~\cite{ziwen1, ziwen2}. In this paper, we propose a measure of the degree of scrambling that employs the entanglement spectrum (ES) statistics. We shall mainly focus on dynamics generated by random unitary circuits without additional conserved quantities, although we do present an example of Hamiltonian dynamics as well. Random unitary circuits serve as an excellent theoretical playground in recent studies of quantum chaos, from which lots of insights on deterministic dynamics can be obtained~\cite{nahum, nahum2, khemani2, rakovszky}. For chaotic random circuits, one expects that the long time evolution \textit{samples uniformly} from the ensemble of Haar-random states~\cite{znidaric1, znidaric2}. Therefore, the entanglement level spacing distribution of the final states of a sufficiently long evolution should be that of Haar-random states, which are described by random matrix theory, in particular, the Gaussian unitary ensemble (GUE) in the present case~\cite{yang2, geraedts, chen}. Crucially, this observation will allow us to define a \textit{distance} between the entanglement level spacing distribution of unitarily evolved states at \textit{intermediate} times --- when the entanglement entropy is far from reaching its maximum --- and the GUE distribution. In this work, we choose as the measure of distance the Kullback-Leibler (KL) divergence, or relative entropy, defined as
\begin{equation}
D_{\rm KL} (p||q) = \sum_i p_i \ {\rm ln} \ \frac{p_i}{q_i} \,,
\end{equation}
and satisfying $D_{\rm KL} \geq 0$, where $\{p_i\}$ and $\{q_i\}$ are two sets of discrete probability distributions {\bf $p,q$}. To compare the degree of randomness that different systems produce under time evolution, we look at $D_{\rm KL}$ as a function of the Page limit-normalized entanglement entropy $S/S_{\rm max}$. This allows for an umambiguous comparison of the degree of randomization between systems under drastically different unitary dynamics beyond the entanglement entropy.

The entanglement level spacing distribution reveals important information regarding the complexity of entanglement that is not captured by the entanglement entropy alone. More precisely, it signals whether a time-evolved state (even if maximally entangled) can be efficiently disentangled \textit{without precise knowledge of the time evolution operator}, which is a highly non-trivial task for generic highly entangled states~\cite{shaffer, chamon, yang2}. Therefore, the distance of the entanglement level distribution to the GUE distribution as a function of the amount of entanglement entropy
should be viewed as the distance to the fixed point under chaotic quantum dynamics, that is, the \textit{degree of scrambling}. In fact, we show in this work that $D_{\rm KL}$ can vary considerably between different chaotic systems \textit{even at the same level of entanglement entropy}.

We demonstrate our results by studying dynamics generated by various random unitary circuits that are chaotic, starting from unentangled product states. Concretely, we investigate two types of random circuits operating on non-Abelian anyons (Fig.~\ref{fig:braid_circuit}): (1) Majorana fermions with random braidings supplemented with random four-body interactions on every four contiguous sites; (2) Fibonacci anyons with random braidings. It is well-known that braidings of Fibonacci anyons alone are capable of universal quantum computation, but braidings of Majorana fermions are not~\cite{freedman, nayak}. In the latter case, in order for the final states to reach GUE entanglement spectra, we
supplement braidings with random \textit{local} four-Majorana interactions. To gain further insights, we also compare the degree of scrambling of anyon braindings with those of the Sachdev-Ye-Kitaev model of Majorana fermions with all-to-all interactions~\cite{sachdev, kitaev2}, and random unitary circuits built out of (a) Hadamard (H), $\pi/8$ (T), and CNOT gates, and (b) random unitary circuits built out of two-qubit Haar-random unitaries.  We find that, at intermediate times (when the entanglement entropy is still far from maximum), the $D_{\rm KL}$ for the above systems are drastically different at the same normalized entanglement entropy $S/S_{\rm max}$. This indicates that there is indeed a hierarchy in the degree of randomness produced by different systems at the same amount of entanglement entropy. Interestingly, we find that braiding of Fibonacci anyons randomizes initial product states more efficiently than the universal H+T+CNOT set, which sheds new light on the potential computational power of topological quantum computation. The SYK model, on the other hand, randomizes more efficiently comparing with local random circuit models.

The rest of this paper is organized as follows. We first introduce the concept of ES statistics and its relevance to entanglement complexity in both quantum circuits and Hamiltonian dynamics in Sec.~\ref{sec:spectrum}. In Sec.~\ref{sec:model}, we introduce the models that we study in this work. The numerical results for $D_{\rm KL}$ are presented in Sec.~\ref{sec:numerics}. Finally we close with a few remarks regarding future directions (Sec.~\ref{sec:summary}).

\section{Entanglement spectrum statistics and entanglement complexity}
\label{sec:spectrum}

We start by introducing the basic concepts of ES statistics and the significance of GUE level spacing statistics in both quantum circuits and Hamiltonian dynamics. 

Consider a pure state $|\psi\rangle$ written in some complete local basis $|\psi \rangle =\sum_{\{\sigma \} } \psi(\{ \sigma \}) |\{\sigma \}\rangle$. From now on we shall drop the ``\{ \}'' and denote a collection of degrees of freedom simply as $\sigma$, which should be clear from the context. Under a bipartition of the system into two subsystems $A$ and $B$ with Hilbert space dimensions $d_A$ and $d_B$, we have:
\begin{eqnarray}
|\psi \rangle &=& \sum_\sigma \psi(\sigma) |\sigma \rangle   \nonumber \\
&=& \sum_{\sigma_A, \sigma_B} \Psi(\sigma_A, \sigma_B) |\sigma_A \rangle |\sigma_B \rangle,
\end{eqnarray}
where in the second line we have recasted the wavefunction $\psi(\sigma)$ as a $d_A \times d_B$ matrix $\Psi(\sigma_A, \sigma_B)$. We define the ES of $|\psi \rangle$ under this bipartition as the set of singular values $\{ \lambda_k \}$ obtained from a Schmidt decomposition of the matrix $\Psi(\sigma_A, \sigma_B)$~\cite{es}. The entanglement entropy can then be defined using the ES as
\begin{equation}
S = -\sum_k \lambda_k^2 \ {\rm ln}\ \lambda_k^2 \,.
\end{equation}
Notice that the reduced density matrix of subsystem $A$ is related to $\Psi$ as $\rho_A = {\rm tr}_B |\psi \rangle \langle \psi | = \Psi \Psi^\dagger$, so the eigenvalues of $\rho_A$ are simply related to the ES as $\{ p_k = \lambda_k^2\}$. Depending on how one partitions the system, $\Psi$ is not necessarily a square matrix. However, in this paper we restrict ourselves to equi-bipartitions, so that $d_A = d_B$.

Historically, the usefulness of the ES was first recognized in the study of \textit{ground states of gapped systems}, and was proposed as a fingerprint of topological order, and even more generally, symmetry breaking order~\cite{li, didier, metlitski, vincenzo, james, vincenzo2, kolley, frank, chandran}. The entanglement entropy of ground states of gapped local Hamiltonians is typically low (area law), which means that the ES decays very fast, usually with a large gap separating the dominant singular values, which capture universal aspects of the underlying system, from the rest of the spectrum. Highly excited states, on the other hand, typically have high entanglement entropy (volume law), and in general the ES is not gapped~\cite{mbl}.

Important physical characteristics encoded in the ES of highly excited states can be revealed in entanglement level spacing statistics. Let the singular values of an ES be rank-ordered in descending order: $\lambda_i > \lambda_{i+1}$; define the ratio of adjacent gaps in the spectrum as
\begin{equation}
r_k = \frac{\lambda_{k-1} - \lambda_k}{\lambda_k - \lambda_{k+1}},
\label{eq:r}
\end{equation}
so that $r_k \geq 0$. It was shown~\cite{chamon, yang, geraedts, yang2, chen} that for Haar-random states, the ES can be described in terms of random matrix theory, where the density of states follows the Marchenko-Pastur distribution~\cite{marchenko}, and the level spacing statistics is given by the Wigner-Dyson distribution~\cite{atas}
\begin{equation}
P_{\rm WD} (r) = \frac{1}{Z} \frac{(r+r^2)^\beta}{(1+r+r^2)^{1+3\beta/2}},
\end{equation}
with $Z=4\pi/81\sqrt{3}$ and $\beta=2$ for GUE distribution. For Poisson distributed spectra, on the other hand, the distribution function takes the form
\begin{equation}
P_{\rm Poisson} = \frac{1}{(1+r)^2},
\end{equation}
which displays no level repulsion as $r\rightarrow 0$ and decays as a different power compared to the GUE distribution as $r \rightarrow \infty$. The ratio~\eqref{eq:r} probes local (nearest-neighbor) correlations between the singular values in the ES. There exists complementary quantities such as the spectral form factor~\cite{cotler,chen} and spectral rigidity~\cite{chamon} which can probe level repulsion at longer ranges. However we will not study these quantities in this work.

The ES level statistics~\eqref{eq:r} contains information on the ``complexity'' of a state, that is, states with GUE distributed ES have complex structures of entanglement, whereas states with Poisson distributed ES have simple structures of entanglement, regardless of the \textit{amount} of entanglement. We define the entanglement complexity as the
inefficiency of \textit{disentangling} a given state evolved under certain unitary dynamics, \textit{without precise knowledge of the time evolution operator}. In general, this is a highly non-trivial task for generic highly entangled states.
However, in Refs.~\cite{chamon, shaffer}, it was shown that one can efficiently disentangle states generated by certain classes of random unitary circuits that are not capable of universal quantum computation (e.g., the Clifford circuits) using a simple Metropolis-like algorithm, \textit{even though these states also reach maximal entanglement entropy}. On the other hand, for states evolved under circuits that are capable of universal quantum computation, such a disentangling algorithm fails. The entanglement entropy growth shows identical behavior in both cases, yet the ES exhibits Poisson distribution in the first case and GUE distribution in the second. Later, Ref.~\cite{yang2} extended this ES-based diagnostic to distinguish between Hamiltonian dynamics that are integrable or non-integrable (either many-body localized or thermalized).

These observations suggest that, in addition to reaching maximal entanglement entropy, time-evolved states of truly chaotic systems should have a GUE distributed ES after a sufficiently long time evolution, i.e., the GUE distributed ES is the fixed point under time evolution, when the initial product states are completely randomized. This motivates us to define the KL divergence between the ES in the process of time evolution and GUE distribution $D_{\rm KL}[P(r)||P_{\rm GUE}(r)]$ as a measure of the degree of randomness produced by the evolution. In particular, when compared at the same fraction of maximal entropy $S/S_{\rm max}$, the difference in $D_{\rm KL}$ reveals the hierarchy of the complexity of scrambling beyond entanglement entropy.

\section{non-Abelian random circuit models}
\label{sec:model}

In this section, we describe the non-Abelian random circuit models that we use for numerical demonstrations of our results. The basic setup we adopt here is to start from unentangled product states, and then evolve under certain types of unitary dynamics. Random unitary circuits have been intensively studied recently as a theoretical handle to understand quantum chaos~\cite{nahum, nahum2, khemani2, von, rakovszky}. In this work, 
we will mainly focus on systems of non-Abelian anyons, with braidings and local interactions acting as unitaries operating on anyonic qubits. These systems provide insights into the degree of scrambling in a context that is also relevant to topological quantum computation.

\subsection{Majorana fermions with random braidings and local interactions}

The simplest non-Abelian anyons carrying a multidimensional representation of the braid group are Majorana fermions, or Ising anyons~\cite{nayak, nayak2}. These are quasiparticle excitations believed to exist in $\nu=5/2$ fractional quantum Hall systems~\cite{moore}, as well as vortex cores of $p+ip$ topological superconductors~\cite{read}. Consider $2n$ Majorana fermions $\gamma_i$ $(i=1,2,\ldots, 2n)$ satisfying $\gamma_i^\dagger = \gamma_i$ and $\{\gamma_i, \gamma_j \} =2\delta_{ij}$. These can be combined into $n$ complex fermions, such that two Majorana fermions fuse into $i\gamma_{2k-1} \gamma_{2k} =1-2n_k=\pm1$, where $n_k$ is the fermion occupation number. Hence the total Hilbert space dimension is $2^{n-1}$ within each fermion parity sector.

\begin{figure}[htb]
\centering
\includegraphics[width=.2\textwidth]{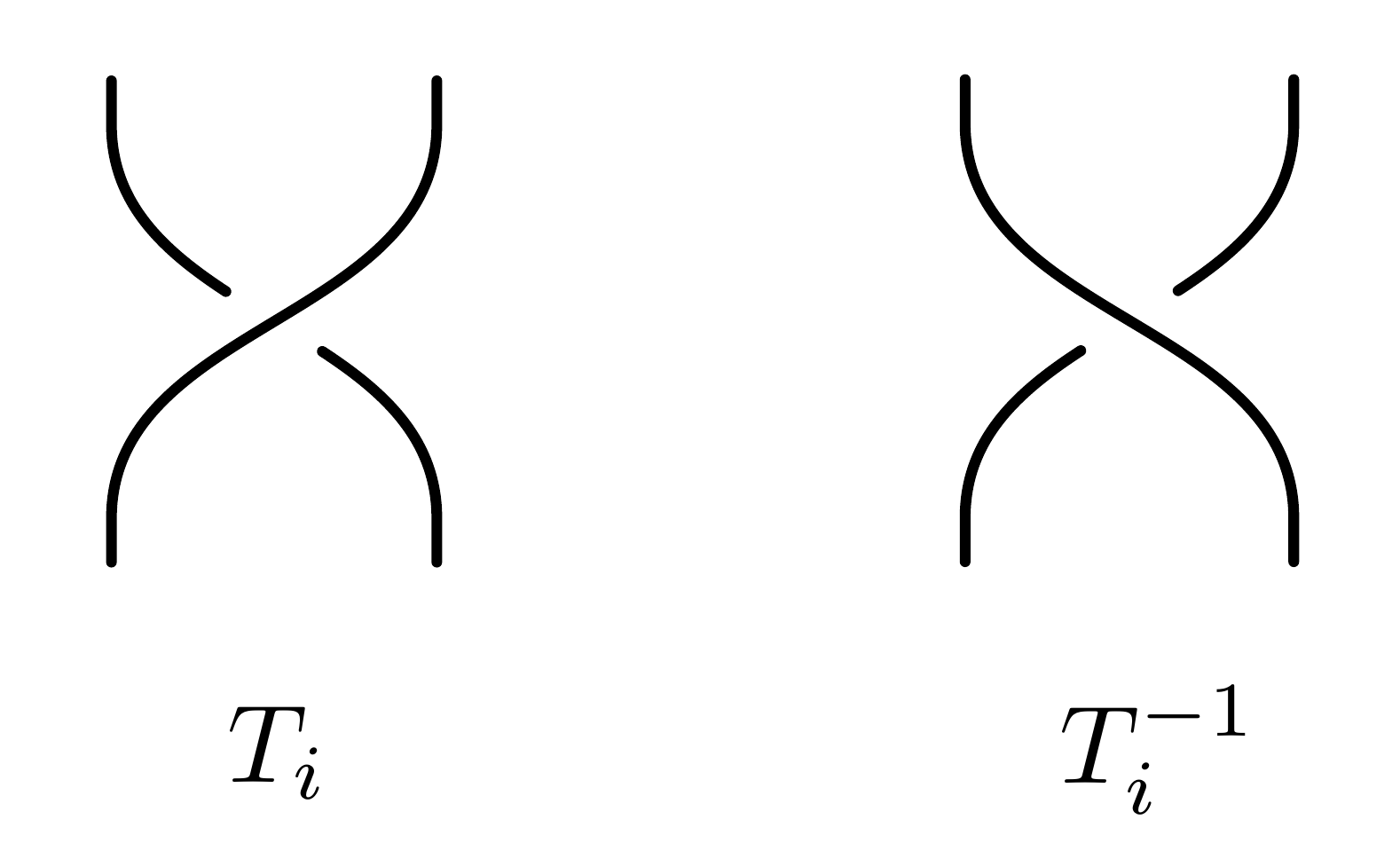}
\caption{(Color online) Generator of the braid group $T_i$ and its inverse $T_i^{-1}$.}
\label{fig:braid_element} 
\end{figure}

Unitary evolutions or ``computations'' are implemented by adiabatically braiding anyons around one another, which induces a transformation on the Hilbert space corresponding to an element of the braid group $B_{2n}$~\cite{kauffman}.  The group $B_{2n}$ is generated by elementary interchanges of neighboring anyons which we denote by $T_i$ (see Fig.~\ref{fig:braid_element}), which satisfy the following relations:
\begin{eqnarray}
T_i T_j &=& T_j T_i, \ \ \ |i-j|>1, \nonumber \\
T_i T_j T_i &=& T_j T_i T_j, \ \ \ |i-j|=1.
\end{eqnarray}
A nonlocal exchange can be achieved via a sequence of nearest-neighbor exchanges; namely, for exchanging anyons $p$ and $q$, one has $T_{p,q} = T_{q-1}T_{q-2}\cdots T_{p+1} T_p T_{p+1}^{-1} T_{p+2}^{-1} \cdots T_{q-1}^{-1}$.

\begin{figure}[htb]
\centering
\includegraphics[width=.45\textwidth]{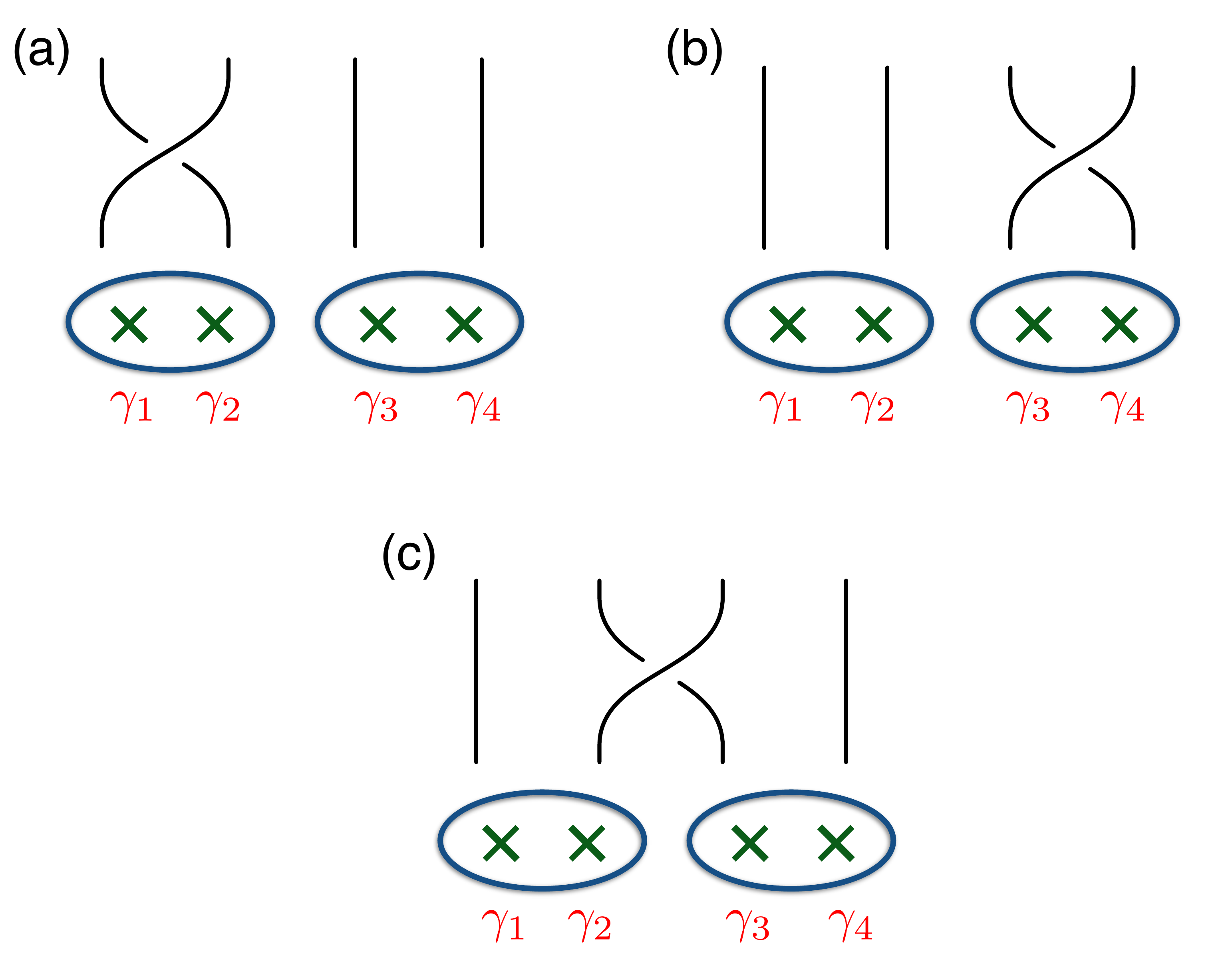}
\caption{(Color online) All possible braid elements acting on four Majorana fermions. (a) $\rho(T_1)$; (b)$\rho(T_3)$; (c)$\rho(T_2)$. The ovals indicate how qubit basis is defined. (a) \& (b) act as single-qubit gates, and (c) acts as a two-qubit gate.}
\label{fig:braid_gate} 
\end{figure}

Physically, we are interested in unitary representations of the braid group. For the case of Majorana fermions, the braid group representation for the generators is given by~\cite{ivanov}:
\begin{equation}
\rho(T_i) = {\rm exp} \left (\frac{\pi}{4} \gamma_{i+1} \gamma_{i} \right) = \frac{1}{\sqrt{2}}(1+\gamma_{i+1} \gamma_i).
\label{eq:braid_Maj}
\end{equation}
Written explicitly under qubit basis, the braid element can act as either a single-qubit gate or a two-qubit gate, depending on whether the two anyons that are braided belong to the same qubit or not. For example, consider four Majorana fermions $\gamma_1, \ldots, \gamma_4$ defining a four-dimensional Hilbert space $|n_1, n_2 \rangle$. One can work out the action of all possible braids on this Hilbert space as given by (see Fig.~\ref{fig:braid_gate}):
\begin{eqnarray}
\rho(T_1)|n_1, n_2 \rangle &=& e^{i\frac{\pi}{4}(1-2n_1)} |n_1, n_2 \rangle,  \nonumber  \\
\rho(T_3)|n_1, n_2 \rangle &=& e^{i\frac{\pi}{4}(1-2n_2)} |n_1, n_2 \rangle,  \nonumber  \\
\rho(T_2)|n_1, n_2 \rangle &=& \frac{1}{\sqrt{2}} \left(|n_1, n_2\rangle + i|1-n_1, 1-n_2 \rangle \right),
\label{eq:braid_Maj_qubit}
\end{eqnarray}
where the braids in the first two lines correspond to single-qubit gates, and the third line corresponds to a two-qubit gate. Applying Eq.~(\ref{eq:braid_Maj_qubit}) to $2n$ Majorana fermions, one can calculate the unitary transformation under circuits of arbitrary braidings of $2n$ anyons.
\\

Braidings of Majorana fermions are insufficient to create circuits capable of universal quantum computation, which is necessary to fully randomize arbitrary initial states. In fact, the braiding representation presented in Eq.~(\ref{eq:braid_Maj}) essentially corresponds to a free fermion system, which fails to even maximally entangle initial product states.
Therefore, we must supplement braidings with local interactions in order to have truly chaotic random circuits~\cite{bravyi}. We add random local four-body interaction terms involving every four contiguous Majorana fermions $\gamma_j, \ldots, \gamma_{j+3}$, which corresponds to unitary operators
\begin{equation} 
U_j = {\rm exp}\left(-i\alpha_j \gamma_j \gamma_{j+1} \gamma_{j+2} \gamma_{j+3}\right),
\label{eq:Maj_inter}
\end{equation}
where $\alpha_j \in [0, 2\pi]$ are random interaction strengths. Now each realization of the random unitary circuit operating on $2n$ Majorana fermions can be built by acting on the anyons with either braiding [Eq.~(\ref{eq:braid_Maj_qubit})] or four-body interaction [Eq.~(\ref{eq:Maj_inter})] at every single step. 

\subsection{Fibonacci anyons with random braidings}
A particular type of non-Abelian anyon that allows for universal quantum computation, and is thus capable of fully randomizing arbitrary initial states, is the Fibonacci anyon~\cite{freedman, bonesteel, hormozi}. The Fibonacci anyon belongs to the quasiparticle spectrum of $SU(2)_3$ Chern-Simons theory whose non-Abelian part is also equivalent to the $\mathbb{Z}_3$ parafermion theory~\cite{read2}. It may also be related to the $\nu=12/5$ fractional quantum Hall state observed in experiments.

The quasiparticle contents in this model are very simple: it contains a single nontrivial quasiparticle denoted as $\tau$ and the identity, or vacuum, denoted by {\bf 1}. The nontrivial fusion rule is given by:
\begin{equation}
\tau \times \tau = \bm{1} + \tau.
\label{eq:fusion_fibo}
\end{equation}

\begin{figure}[htb]
\centering
\includegraphics[width=.4\textwidth]{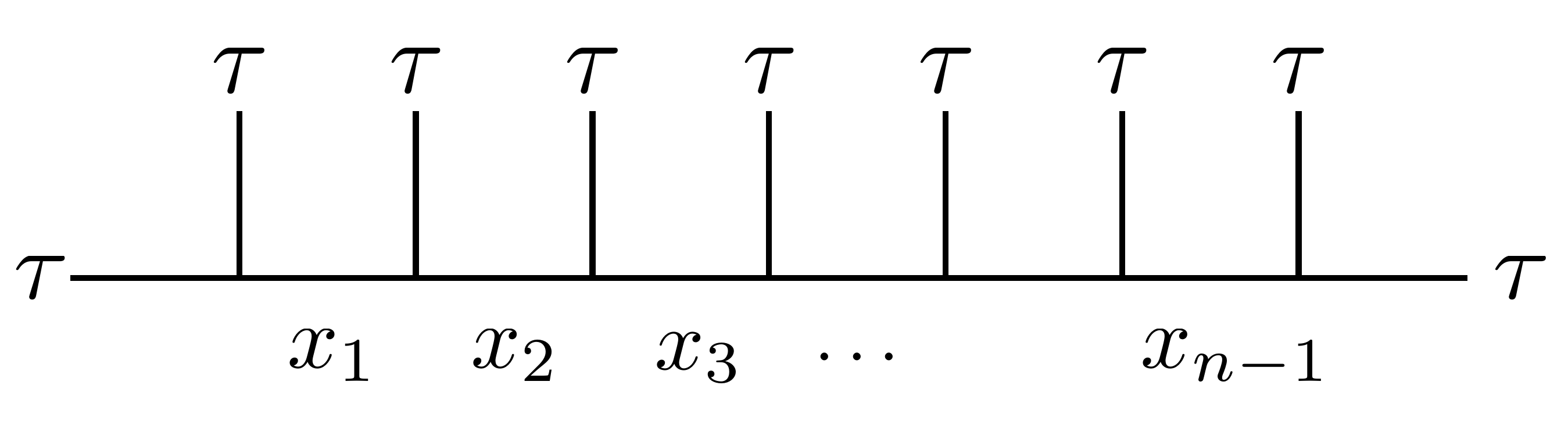}
\caption{(Color online) The fusion tree of a Fibonacci chain consisting of $n$ anyons. States in the Hilbert space are labeled by the degrees of freedom on the horizontal links $|x_1x_2\ldots x_{n-1} \rangle$, with the additional constraint that there cannot be two {\bf 1}'s next to each other.}
\label{fig:fibo_chain} 
\end{figure}

To define the Hilbert space of a system of multiple anyons, we consider the Fibonacci chain~\cite{feiguin, chandran2} with open boundary condition as shown in Fig.~\ref{fig:fibo_chain}. This is essentially a fusion tree drawn in a slightly different orientation.
We label states in the Hilbert space corresponding to Fig.~\ref{fig:fibo_chain} as $|x_1x_2\ldots x_{n-1} \rangle$, with $x_i=\tau$ or {\bf1}, for a system of $n$ anyons (the $\tau$'s on the two endpoints are considered as boundary conditions). Due to the fusion rule in Eq.~(\ref{eq:fusion_fibo}), the allowed configuration must satisfy an additional constraint, namely, there cannot be two {\bf 1}'s next to each other. Hence the Hilbert space dimension of $n$ anyons is Fib($n+1$), where Fib($n$) are Fibonacci numbers satisfying Fib($n+1$) = Fib($n$) + Fib($n-1$) and Fib(1) = Fib(2) = 1. For large $n$, the Hilbert space dimension of Fibonacci anyons grows as $\phi^n$, where the quantum dimension $\phi = (1+\sqrt{5})/2$ is the golden ratio.

\begin{figure}[htb]
\centering
\includegraphics[width=.2\textwidth]{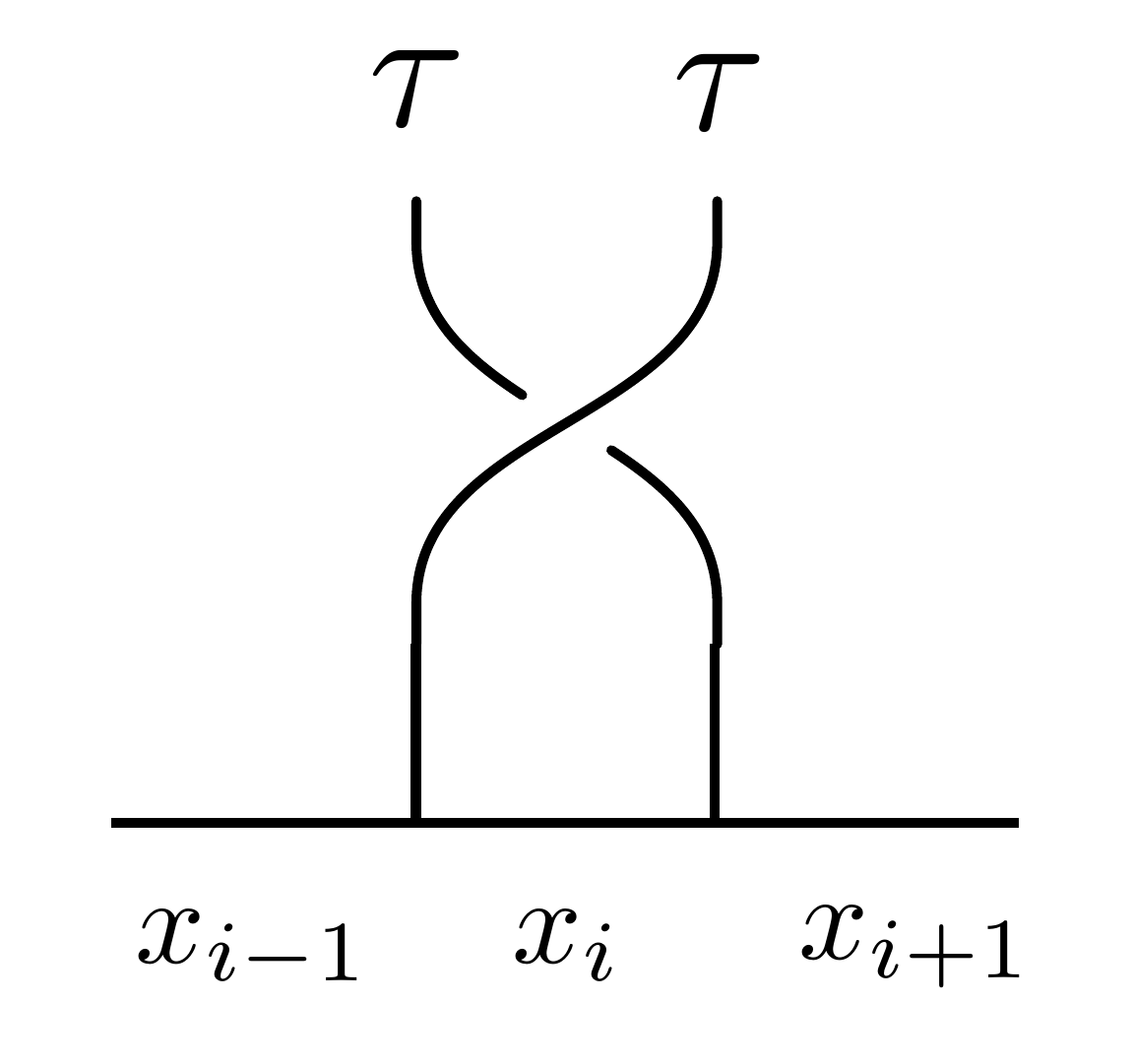}
\caption{(Color online) The effect of braiding two Fibonacci anyons only depends on the configuration of the three qubits in contact with the two anyons: $x_{i-1},\ x_i$, and $x_{i+1}$.}
\label{fig:gate_braid_fibo} 
\end{figure}

The unitary representation of the braid group in terms of Fibonacci anyons can be derived using the $R$-matrix and the $F$-matrix of the topological field theory. From now on we shall denote the states of $x_i$ using the language of qubits: $|0\rangle \equiv |x_i={\bm 1}\rangle$, $|1\rangle \equiv |x_i = \tau \rangle$. We directly list below the action of braiding on the qubit basis as in Eq.~(\ref{eq:braid_Maj_qubit}) and provide the derivation in Appendix A. The effect of braiding two adjacent anyons depends only on the configuration of the three qubits in contact with the two anyons, see Fig.~\ref{fig:gate_braid_fibo}. The representation of the braiding in the qubit basis is
\begin{eqnarray}
\rho(T_i) \ |101\rangle &=& -e^{-i\pi/5}/\phi \ |101\rangle -i e^{-i\pi/10}/\sqrt{\phi} \ |111\rangle, \nonumber \\
\rho(T_i) \ |111\rangle &=& -ie^{-i\pi/10}/\sqrt{\phi}\ |101\rangle - 1/\phi \ |111\rangle,  \nonumber  \\
\rho(T_i) \ |110\rangle &=& -e^{-2\pi i/5} \ |110\rangle,  \nonumber  \\
\rho(T_i) \ |011\rangle &=& -e^{-2\pi i/5} \ |011\rangle,  \nonumber  \\
\rho(T_i) \ |010\rangle &=& e^{-4\pi i/5} \ |010\rangle,
\label{eq:braid_fibo_qubit}
\end{eqnarray}
where we have suppressed the labels for the rest of the qubits.

The ``no consecutive \textbf{1}s'' constraint imposed on the states of adjacent qubits means that the Hilbert space dimension of the full chain is \textit{not} equal to the product of the subsystem Hilbert space dimensions under a bipartition, i.e. $d \neq d_A d_B$. 

\subsection{Hadamard, $\pi/8$, and CNOT gate}

In addition to the two non-Abelian random circuit models introduced above, we also study three more cases as a comparison. The first system is the random circuits built out of Hadamard, $\pi/8$, and CNOT gates. The action of these gates is most conveniently expressed in terms of the following unitary matrices:
\begin{eqnarray}
H = \frac{1}{\sqrt{2}} 
\begin{pmatrix}
1 & 1 \\
1 & -1
\end{pmatrix},
&&\ \ \ T =
\begin{pmatrix}
1 & 0 \\
0 & e^{i\pi/4}
\end{pmatrix}, \nonumber  \\
{\rm CNOT} &=&
\begin{pmatrix}
1 & 0 & 0 & 0 \\
0 & 1 & 0 & 0 \\
0 & 0 & 0 & 1 \\
0 & 0 & 1 & 0
\end{pmatrix}.
\end{eqnarray}
It can be shown that H, T and CNOT combined together is capable of universal quantum computation~\cite{nielsen}, and hence is also capable of fully randomizing arbitrary initial states.

\subsection{Two-qubit Haar-random unitaries}
In this case, we study the same setup as in previous work by considering random unitary circuits built from two-qubit Haar-random unitary gates, which are drawn from the uniform probability distribution on the unitary group for two qubits $U(4)$~\cite{nahum, nahum2}. Here we shall only use local gates which act on two nearest-neighboring qubits at a single step.

\subsection{SYK model}
Finally, we consider the SYK model consisting of $N$ Majorana fermions with all-to-all random interactions~\cite{sachdev, kitaev2}. The main difference from all previous cases is that this system undergoes Hamiltonian dynamics with energy conservation, instead of random unitary dynamics. The Hamiltonian of the SYK model is:
\begin{equation}
\mathcal{H} = \sum_{ijkl} J_{ijkl} \gamma_i \gamma_j \gamma_k \gamma_l,
\label{eq:syk}
\end{equation}
where the couplings $J_{ijkl}$ are real Gaussian random variables with zero mean and variance $\overline{J_{ijkl}^2}=3!J^2/N^3$. This model is exactly solvable in the large-$N$ limit, with a Lyapunov exponent obtained from the OTOC saturating the bound $\lambda_L = 2\pi/\beta$, where $\beta$ is the inverse temperature~\cite{maldacena, maldacena2}. Hence the SYK model is often referred to as being maximally chaotic. Moreover, the eigenstates of the SYK model have been shown to be thermalizing, with entanglement entropies obeying volume law scaling~\cite{fu, sonner, liu, huang}.
It is thus interesting to look at the degree of scrambling
using our measure of $D_{\rm KL}$
and compare with the previous cases.

We consider pure states obtained from a \textit{quantum quench} of an unentangled product state with the Hamiltonian of Eq.~(\ref{eq:syk})~\cite{kourkoulou, eberlein}.
One can explicitly construct a representation for the Majorana field operators in terms of Pauli matrices using the following Jordan-Wigner transformation:
\begin{equation}
\gamma_{2k-1} = \sigma_k^x \prod_{i=1}^{k-1} \sigma_i^z,  \ \ \ \ \gamma_{2k} = \sigma_k^y \prod_{i=1}^{k-1} \sigma_i^z,
\end{equation}
which maps the Hamiltonian~(\ref{eq:syk}) to a spin system.

\section{numerical results}
\label{sec:numerics}
We now present our numerical calculations of $D_{KL}$ as a function of $S/S_{\rm max}$ for models A-E as explained in previous sections. Here the $S_{\rm max}$ denotes the maximal Page entropy of a Haar-random state given by~\cite{page}:
\begin{equation}
S_{\rm max} = {\rm ln}d_A - \frac{d_A}{2d_B},~~~d_A\leq d_B
\end{equation}
where $d_A=d_B = $ Fib($n_A+2$) for the Fibonacci anyon model with $n_A$ qubits (not anyon number!) in subsystem $A$, and $d_A=d_B=2^{n_A}$ for all other cases~\cite{fib}.

For both the Majorana fermion circuit and the SYK model, the global fermion parity is conserved, and physically one can only create states within a fixed fermion parity sector: $i^{n} \gamma_1\gamma_2\cdots \gamma_{2n} |\psi\rangle = \pm |\psi \rangle$. As a consequence of that, the reduced density matrix $\rho_A$ will be block-diagonal with two blocks corresponding to even/odd parities respectively, and hence the ES statistics of the full spectrum will be a mixture of two sectors and yields a Poisson distribution. One way of working around this is to study the ES statistics within each block. However, in order to obtain a denser spectrum without having to double the total number of sites, a more convenient way is to simply start with random product states that mix the two parity sectors. The physics of scrambling should not be affected by this choice, and one should view this as a theoretical probe, not the modeling of a physical system.

\begin{figure}[t]
\centering
\includegraphics[width=.47\textwidth]{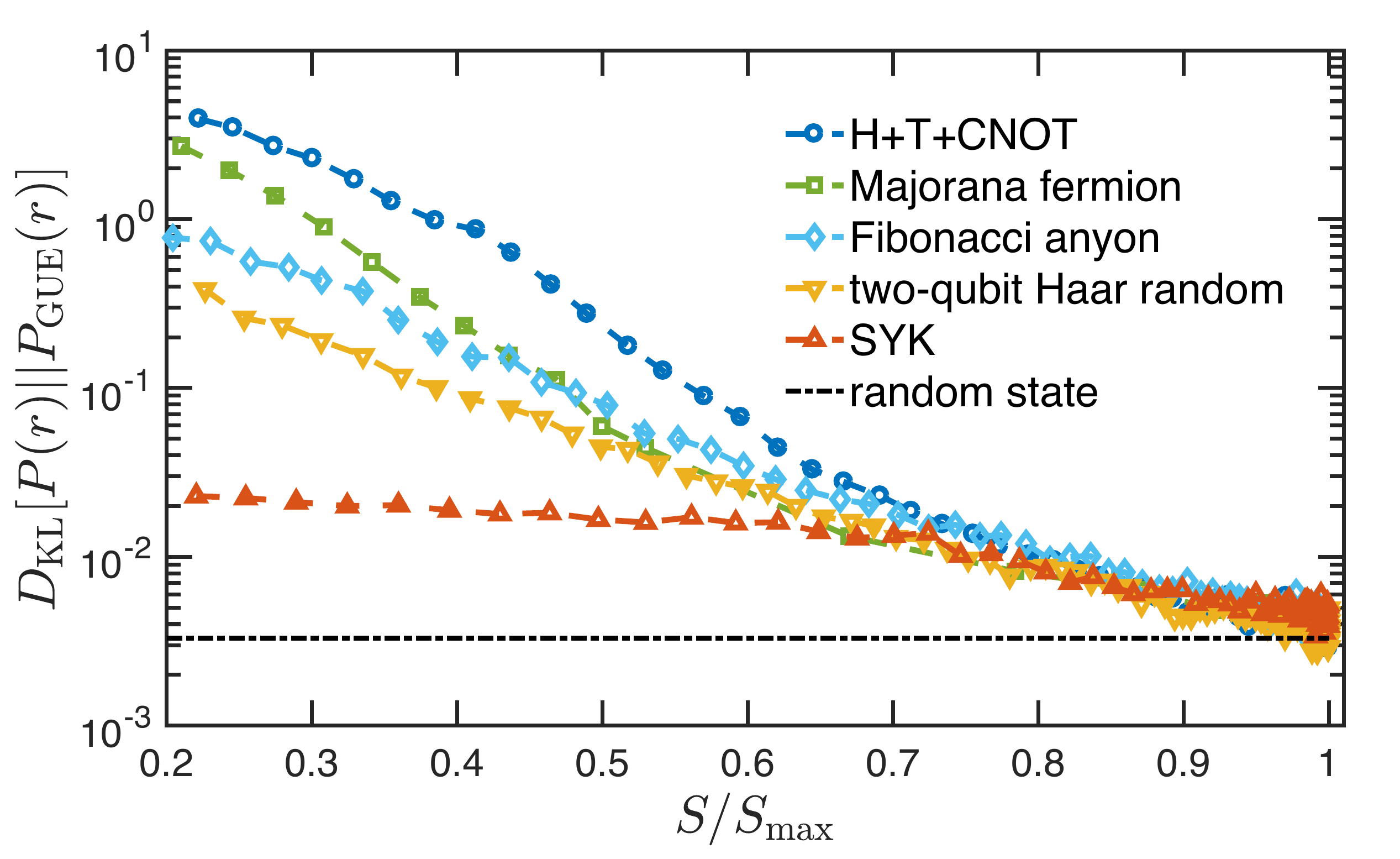}
\caption{(Color online) $D_{\rm KL}$ as a function of $S/S_{\rm max}$ for different models. At intermediate times, one clearly observes a hierarchy of $D_{\rm KL}$ among various chaotic systems \textit{even at the same amount of entropy}. The horizontal dotted dashed line corresponds to the $D_{\rm KL}$ calculated for Haar-random states, which serves as a lower bound numerically. We only look at times after $S/S_{\rm max}>0.2$ when we have enough non-zero singular values in each ES to study statistics. For each ES, singular values smaller than $10^{-12}$ are discarded. The data are obtained for: $N=28$ Majorana fermions (equivalently 14 qubits) for the SYK model, averaged over 2000 realizations; $n=23$ anyons (equivalently 22 qubits) for the Fibonacci anyon model, averaged over 1000 realizations; and 16 qubits averaged over 1000 realizations for all other cases.}
\label{fig:dkl} 
\end{figure}

The numerical results are summarized in Fig.~\ref{fig:dkl}. Remarkably, we find that there is a clear hierarchy of $D_{\rm KL}$ among various models \textit{even at the same amount of entropy}. Since $D_{\rm KL}$ quantifies how close a given state is to a Haar-random state, this result reveals the hierarchy in the degree of randomness produced by different unitary evolutions beyond the entanglement entropy. As shown in Fig.~\ref{fig:dkl}, at intermediate times (when the entanglement entropy is still far from maximum), the H+T+CNOT gate set turns out to be the least efficient scrambler among all cases. This result shows that, although the universal set of gates is capable of scrambling initial product states, at intermediate steps, the degree of randomness of such states are in fact quite low. In particular, in this regime, the states generated by H+T+CNOT gates are much less random than those generated by two-qubit Haar-random unitary circuits, at the same amount of entanglement entropy.

Interestingly, the Fibonacci anyon model randomizes initial product states more efficiently than the H+T+CNOT gate set. Traditionally, there has been considerable effort in designing sophisticated compiling algorithms to build H, T, and CNOT gates from either the Ising or Fibonacci anyons~\cite{bravyi, bonesteel, hormozi}. While this is necessary for implementations of real-world quantum algorithms written in terms of universal gates, our results imply that a better strategy may be to bypass the compilation of standard gates completely and focus on computations carried out directly with braiding. Conversely, one can ask how to approximate an arbitrary anyonic braid by combining universal gates. Our results indicate that the appropriate combination may be quite complicated.

The SYK model, in spite of having additional energy conservation, generates a much higher degree of randomness than the local random unitary circuit models. This result may not be completely surprising since the SYK model has all-to-all interactions, hence the large number of independent random couplings in the Hamiltonian makes the corresponding unitary resemble a Haar-random unitary acting on $n$ qubits. However, we now have a concrete way to quantify the degree of randomness that this system produces under time evolution, which can be compared with systems under drastically different dynamics.

It is important to emphasize that, by looking at $D_{\rm KL}$ as a function of the normalized entanglement entropy $S/S_{\rm max}$, we are not comparing the `speed' of information scrambling that people usually think of. After all, it is not sensible to talk about a unique time unit for drastically different systems, namely, random unitary circuits versus local or non-local Hamiltonian systems. Rather, we compare the degree of randomness produced by different systems when the same fraction of the systems becomes entangled, that is, at the same $S/S_{\rm max}$. One may alternatively view the parametrization $S/S_{\rm max}$ as a `proper time' which eliminates real time or circuit depths and allows different models to be compared on equal footing. 
In Appendix B, we show that this parametrization is insensitive to different system sizes, whereas the real time or circuit depth is not. 

Finally, there is an interesting observation from Fig.~\ref{fig:dkl} regarding the late time behavior of $D_{\rm KL}$ (when the entanglement entropy is close to reaching its maximum). Starting from $S/S_{\rm max} \approx 0.7$, the curves for different cases seem to collapse on top of each other. It is tempting to think of this as a universal late-time behavior of chaotic systems as a function of the entropy, in the sense that the discrepancies in the degree of randomness they produce disappear prior to reaching maximal entanglement entropy.

\section{Summary and outlook}
\label{sec:summary}
The notion of information scrambling bridges many different areas in physics, including quantum many-body physics, quantum computation and quantum information, quantum statistical mechanics and quantum gravity. Scrambling can exhibit different complexities depending on the degree of randomness it produces, which very often cannot be captured by the entanglement entropy alone~\cite{ziwen1,ziwen2}.
In this work, we propose a new metric to quantify the degree of scrambling in the form of the Kullback-Leibler divergence $D_{\rm KL}$, which is a measure of the distance to universal ES level spacing distribution corresponding to complete randomization. The universal distribution of the ES is intimately tied to the complexity of entanglement in a given state, which is not reflected in the net amount of entanglement entropy. We demonstrate numerically that there is indeed a hierarchy of $D_{\rm KL}$ among various models, which defines the degree of scrambling in a model-independent manner.

This work opens interesting directions for future research. Our methodology can be extended to a plethora of other quantum Hamiltonian systems, such as quantum spin chains. In particular, it was pointed out in Ref.~\cite{yang2} that many-body localized (MBL) systems --- although they do not thermalize and hence do not reach Page entropy --- also reach GUE distributed ES asymptotically as $1/{\rm ln}t$, where $t$ is real time. Therefore, $D_{\rm KL}$ could potentially be a useful quantity to compare the degree of scrambling of different MBL systems, without the ambiguity of time units.

Furthermore, it has been conjectured that quantum chaos underlies the computational complexity of quantum circuits or, more generally, quantum channels~\cite{hosur, cotler}. From the topological quantum computation perspective, our results imply that braiding non-Abelian anyons may well be a much faster quantum computer than the universal gate set. This suggests that there might be more efficient ways of utilizing the computational power of braidings, than trying to design the universal gates using braidings.

\section*{Acknowledgments}
We would like to thank Meng Cheng, Yingfei Gu, and Eric Rowell for useful discussions, and Marko \v{Z}nidari\v{c} for useful comments on the manuscript. We thank Zi-Wen Liu for pointing out refs.~\cite{ziwen1,ziwen2} to us and helpful discussions. Z.-C. Y. and C. C. are supported by DOE Grant No. DE-FG02-06ER46316. S. K. was partially supported by the Boston University Center for Non-Equilibrium Systems and Computation. K. M. acknowledges the Boston University Condensed Matter Theory Visitors Program for its hospitality and financial support from the EPSRC Doctoral Prize Fellowship.

\appendix
\section{Derivation of the braid group representation in terms of Fibonacci anyons}

We present the derivation of the action of braiding operations on qubit basis of Fibonacci anyons, whose results are summarized in Eq.~\ref{eq:braid_fibo_qubit}. We shall use the language of $F$ and $R$ matrices to derive these results. There also exists an alternative derivation by drawing a connection between the braid group representation and the Temperley-Lieb algebra. We refer the interested readers to refs.~\cite{kauffman2, delaney}.

\begin{figure}[htb]
\centering
\includegraphics[width=.4\textwidth]{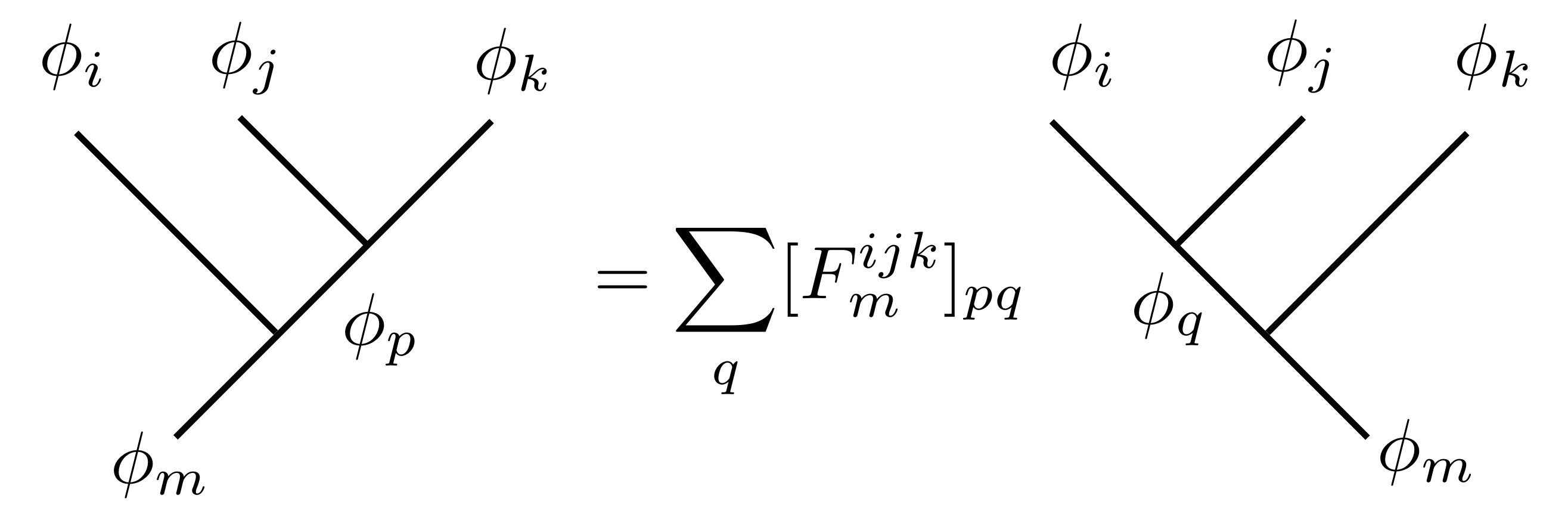}
\caption{(Color online) The $F$-matrix relates the two different basis states resulting from fusing four anyons in different orders.}
\label{fig:fmatrix} 
\end{figure}

The $R$-matrix and $F$-matix are important data characterizing a given conformal field theory or topological quantum field theory. The $R$-matrix $R^{ab}_c$ specifies the phase resulting from braiding anyons of type $a$ and $b$ which fuse into type $c$. For the Fibonacci anyons, the $R$-matrix is given by~\cite{nayak}:
\begin{equation}
R =
\begin{pmatrix}
e^{4\pi i/5} & 0 \\
0 & -e^{2\pi i/5}
\end{pmatrix},
\end{equation}
or, $R^{\tau \tau}_{\bm{1}}=e^{4\pi i/5}$, $R^{\tau \tau}_{\tau} = -e^{2\pi i/5}$. The $F$-matrix $[F^{ijk}_m]_{pq}$, on the other hand, specifies the unitary transformation between two difference bases, when four anyons are fused in different orders (see Fig.~\ref{fig:fmatrix}). 
For Fibonacci anyons, $F^{\tau \tau \tau}_{\bm{1}}$ can be easily seen to be trivially identity. However, $F^{\tau \tau \tau}_{\tau}$ is non-trivial and given by~\cite{nayak}:
\begin{equation}
F^{\tau \tau \tau}_{\tau} =
\begin{pmatrix}
1/\phi & 1/\sqrt{\phi} \\
1/\sqrt{\phi} & -1/\phi
\end{pmatrix},
\end{equation}
where $\phi = (1+\sqrt{5})/2$ as defined earlier. Notice that the $F$-matrix satisfies $F^{-1} = F$.

\begin{figure}[htb]
\centering
\includegraphics[width=.15\textwidth]{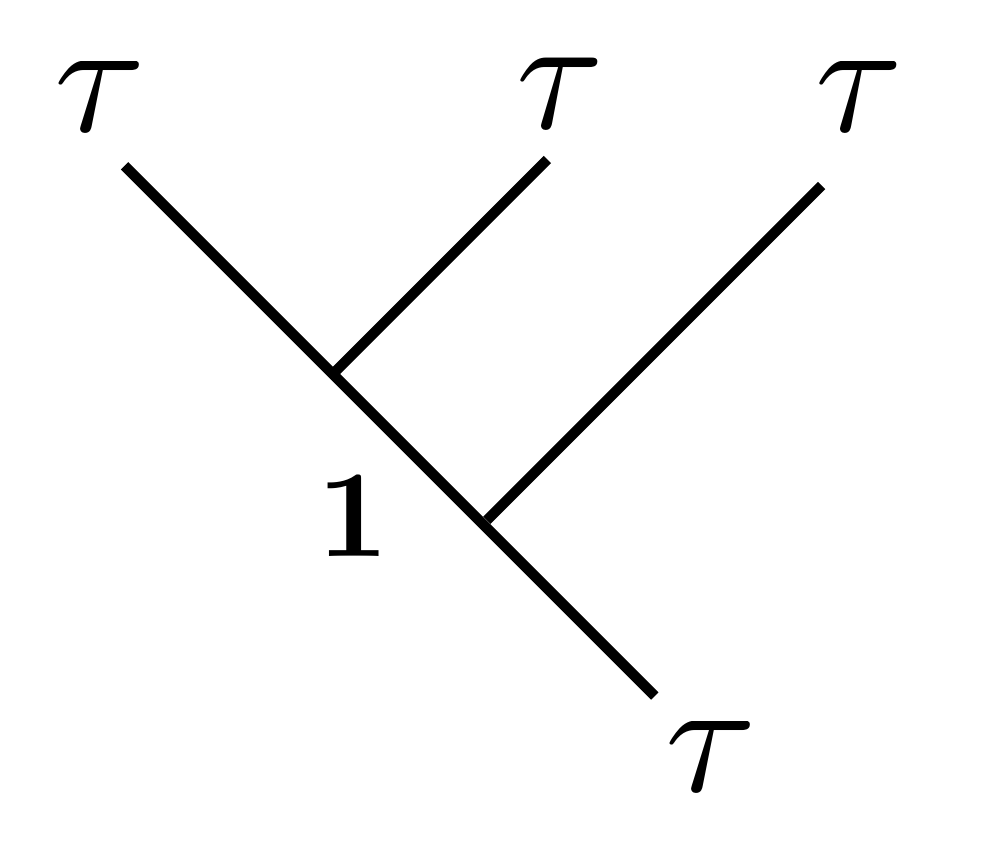}
\caption{(Color online) A qubit configuration $|x_{i-1}x_i x_{i+1}\rangle = |101\rangle$ corresponding to Fig.~\ref{fig:gate_braid_fibo} drawn in the fusion tree orientation.}
\label{fig:braid_fibo_eg} 
\end{figure}

As we explained in Fig.~\ref{fig:gate_braid_fibo}, the action of braiding on the qubit basis only depends on the configuration of three adjacent qubits. One can tilt Fig.~\ref{fig:gate_braid_fibo} and draw it in exactly the same way as in Fig.~\ref{fig:fmatrix}, so that the action of braiding can be determined using the $F$ and $R$ matrices. Let us give an example of the state $|x_{i-1}x_i x_{i+1}\rangle = |101\rangle$. This configuration corresponds to the fusion tree depicted in Fig.~\ref{fig:braid_fibo_eg}.
One will first need to convert to the basis where the two anyons being braid (top right) fuse into a definite anyon type using the (inverse of) $F$-matrix, then followed by applying the $R$-matrix. Finally, one converts back to the original basis by applying $F$ again. Hence, the action of braiding on this configuration is given by the first row of the matrix $FRF^{-1}$, which is:
\begin{equation}
FRF^{-1} =
\begin{pmatrix}
-e^{-i\pi/5}/\phi & -i e^{-i\pi/10}/\sqrt{\phi} \\
-i e^{-i\pi/10}/\sqrt{\phi} & -1/\phi
\end{pmatrix}.
\label{eq:frf}
\end{equation}
Eq.~\ref{eq:frf} leads to the first line of Eq.~\ref{eq:braid_fibo_qubit} shown in the text. One can also check in a similar way the rest of the results claimed in Eq.~\ref{eq:braid_fibo_qubit}.

\section{Finite-size effect on the parametrization $S/S_{\rm max}$}
In this section, we show that the parametrization $S/S_{\rm max}$ is insensitive to finite-size effect, whereas the real time or circuit depth is not. Therefore, the main result that we present in Fig.~\ref{fig:dkl} holds even with the slight non-uniformity in our choices of system sizes for different models. 

We take the case of universal set of gates H+T+CNOT as an example. 
\begin{figure}[htb]
\centering
\includegraphics[width=.45\textwidth]{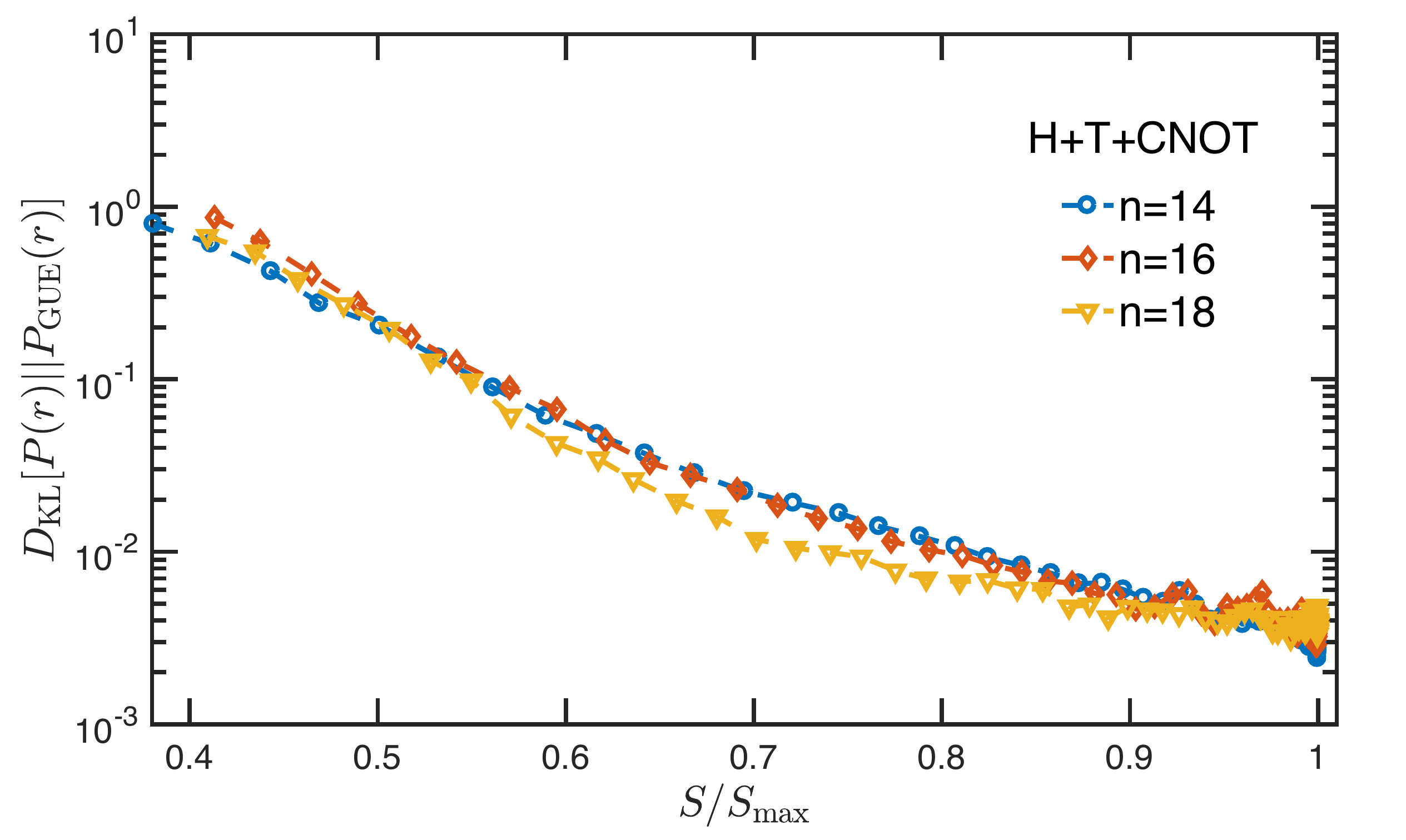}
\centerline{(a)}
\includegraphics[width=.46\textwidth]{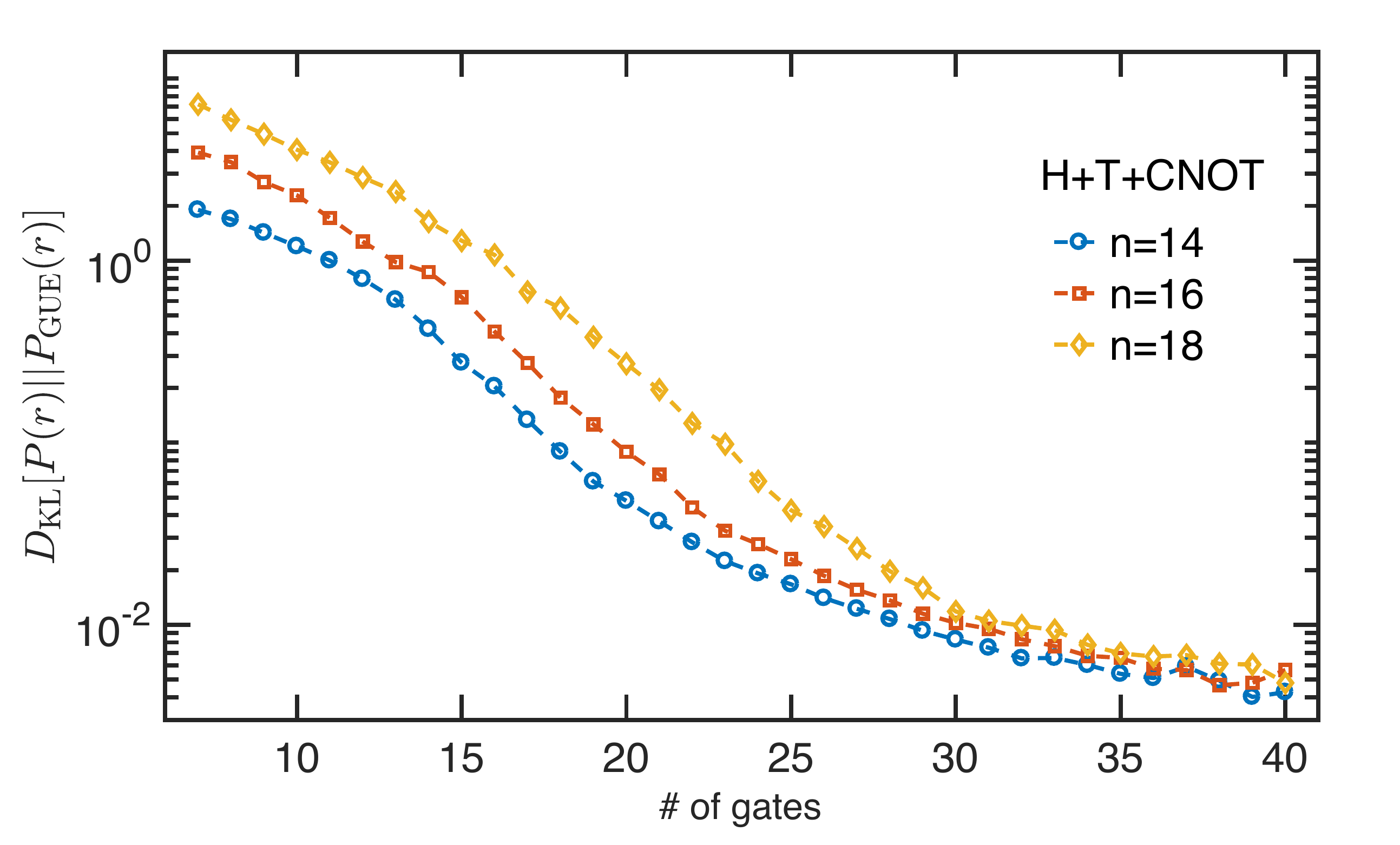}
\centerline{(b)}
\caption{(Color online) (a)$D_{\rm KL}$ as a function of $S/S_{\rm max}$ for the H+T+CNOT gate set for different system sizes; (b)$D_{\rm kL}$ as a function of circuit depth (i.e. real time) for different system sizes. One finds that the curves collapse with the parametrization $S/S_{\rm max}$ but not with the circuit depth.}
\label{fig:finitesize} 
\end{figure}
In Fig.~\ref{fig:finitesize}(a), we plot $D_{\rm KL}$ as a function of $S/S_{\rm max}$ for different system sizes. We find that the curves almost fall on top of one another, indicating that the ratio $S/S_{\rm max}$ measures the fraction of the degrees of freedom of the system that becomes entangled and is thus insensitive to different system sizes. On the other hand, if we plot $D_{\rm KL}$ as a function of the circuit depth (i.e. real time) as in Fig.~\ref{fig:finitesize}(b), we find that there is a systematic shift of the curves upon changing system sizes, namely, $D_{\rm KL}$ gets bigger at the same circuit depth as the system size increases. This is consistent with the expectation that it takes a longer real time for larger systems to scramble to the same degree. However, this effect is indeed eliminated with the parametrization $S/S_{\rm max}$, which can be viewed as a `proper time' that is robust against slight non-uniformity in the system sizes.

\bibliography{manuscript}

\begin{thebibliography}{75}
\bibitem [{es()}]{es}%
  \BibitemOpen
  \href@noop {} {}\bibinfo {note} {There exits an alternative definition of the
  ES that commonly appears in the literature, by writing the reduced density
  matrix in the form $\rho_A = {\rm exp}(-H)$ and defining the eigenvalue
  spectrum of the "entanglement Hamiltonian" $H$ as the ES. For our current
  purpose of level spacing statistics, one can show that these two definitions
  do not make a difference. We shall adopt the definition in the text for
  convenience.}\BibitemShut {Stop}%
\end{thebibliography}

\end{document}